# Entropy-statistical approach to phase-locking detection of pulse oscillations: application for the analysis of biosignal synchronization

Petr BORISKOV[†‡], Vadim PUTROLAYNEN[†], Andrei VELICHKO[†] and Kristina PELTONEN[†]

[†]*Institute of Physics and Technology, Petrozavodsk State University, 185910 Petrozavodsk, Russia*

[‡] Corresponding author

This research was supported by the Russian Science Foundation (grant no. 22-11-00055, https://rscf.ru/en/project/22-11-00055/, accessed on 30 March 2023).

Petr Boriskov, boriskov@petrsu.ru, ORCID: http://orcid.org/0000-0002-2904-9612; Vadim Putrolaynen, vputr@petrsu.ru, ORCID: http://orcid.org/0000-0001-5707-7760; Andrei Velichko, velichkogf@gmail.com, ORCID: http://orcid.org/0000-0002-9341-1831; Kristina Peltonen, krispelt@yandex.ru

**Abstract: In this study a new method for analyzing synchronization in oscillator systems is proposed using the example of modeling the dynamics of a circuit of two resistively coupled pulse oscillators. The dynamic characteristic of synchronization is fuzzy entropy (FuzzyEn) calculated a time series composed of the ratios of the number of pulse periods (subharmonic ratio, SHR) during phase-locking intervals. Low entropy values indicate strong synchronization, whereas high entropy values suggest weak synchronization between the two oscillators. This method effectively visualizes synchronized modes of the circuit using entropy maps of synchronization states. Additionally, a classification of synchronization states is proposed based on the dependencies of FuzzyEn on the length of embedding vectors of SHR time series. An extension of this method for analyzing non-relaxation (non-spike) type signals is illustrated using the example of phase-phase coupling rhythms of local field potential of rat hippocampus. The entropy-statistical approach using rational fractions and pulse signal forms makes this method promising for analyzing biosignal synchronization and implementing the algorithm in mobile digital platforms.**

## Contributors

Petr Boriskov designed the research; Petr Boriskov and Vadim Putrolaynen processed the data; Petr Boriskov, Vadim Putrolaynen, and Andrei Velichko drafted the manuscript; Andrei Velichko and Kristina Peltonen helped organize the manuscript; Vadim Putrolaynen and Andrei Velichko revised and finalized the paper.

## Compliance with ethics guidelines

Petr Boriskov, Vadim Putrolaynen, Andrei Velichko, and Kristina Peltonen declare that they have no conflict of interest.

## 1 Introduction

Synchronization is a fundamental phenomenon that plays a crucial role in various fields of science and technology (Pikovsky Author A et al., 2002; Park J et al., 2003; Belluscio M et al., 2012; Lowet E et al., 2016; Nikonov D E et al., 2015). For example, in all digital communication systems, synchronization is essential for establishing precise phase relationships between the transmission and reception cycles of digital signals (Park J et al., 2003). In neurophysiology, neuronal oscillatory synchronization indicates the presence of connections between neurons or their populations, which is considered the basis for information coordination between cortical neural networks (Belluscio M et al., 2012; Lowet E et al., 2016).

A new direction in neural network technologies, known as oscillator computing, is entirely based on the effect of synchronization (Nikonov D E et al., 2015; Mallick A et al., 2020; Romera M et al., 2018; Chou J et al., 2019; Wang T and Roychowdhury J, 2019). For instance, the ability to perform vowel classification using the frequency synchronization characteristics of four coupled spin-torque nano-oscillators has been demonstrated (Romera M et al., 2018). J. Chou et al. (2019) and T. Wang et al. (2019) solved the Max-Cut problem using synchronization of LC-oscillators.

The regular or irregular nature of dynamics in many applied tasks indicates the "correct" or "incorrect" systems operation. Examples include jitter in digital devices, which is a primary source of distortion in frequency synthesizers (Sotiriadis P P, 2010), anomalies in electrocardiograms (Becker D, 2006), irregularities in the rotational speed of gears and motors (Jiang H et al., 2015; Bonet-Jara J et al., 2021), and the registration of irregularities in earthquake precursors (Biagi P et al., 2001). The regularity of signals is often assessed by calculating their entropy, such as fuzzy (Ishikawa A and Mieno H, 1979), sample (Delgado-Bonal A and Marshak A, 2019), or SVD (Alter O et al., 2000) entropy indicators. Synchronization of oscillations in interacting systems is closely related to the regularity of their joint dynamics. Therefore, an approach to detecting and analyzing oscillation synchronization through entropy assessment is promising and already used in various systems (Jiang H et al., 2015; Bonet-Jara J et al., 2021; Biagi P et al., 2001; Ishikawa A and Mieno H, 1979).

Among the most commonly used characteristics of biosignal synchronization is the phase-locking parameter (PLP) (Pikovsky Author A et al., 2002; Belluscio M et al., 2012; Lowet E et al., 2016):

$$P_{nm} = \frac{1}{T}\left|\sum_{t=1}^{T}\exp\left(i\Delta\varphi_{nm}(t)\right)\right|, \quad (1)$$

where $\Delta\phi_{\mathrm{nm}}(t)$ is the phase difference between signals with integers $n$ and $m$ (see section 2.2), and $T$ is the total number of time points in the sample. Essentially, PLP is a set of integral characteristics for different $n$ and $m$, computed from a large number of samples. The result of these calculations is the classification of the system as likely synchronized or not for a given $n$ and $m$. Another metric of signal synchronization is spectral coherence. This is the most comprehensive assessment of synchronization (Lowet E et al., 2016), but it requires the calculation of spectra and the mutual correlation function of signals, making it as complex as PLP.

In recent studies (Velichko A et al., 2018), we proposed simple metrics for the synchronization of relaxation signals (spikes): a time series consisting of subharmonic ratio (SHR) coefficients, and SHR efficiency (SE), determined from the SHR time series statistics. Based on these metrics, a new type of spike neural networks (Velichko A et al., 2018; Velichko A, 2019) with high-order synchronization (HOS) effects was developed. The SHR algorithm is high-speed and does not require large computing resources.

This paper develops a new approach to analyzing HOS of pulse signals, using fuzzy entropy (FuzzyEn) calculated from the SHR time series as the synchronization characteristic. The major contributions of the paper are as follows:

- A model of a digital-analog circuit of resistively coupled oscillators was constructed, including modes of frequency-modulated noise impact and HOS state modeling.
- An algorithm for calculating the SHR time series using the logical AND function of the oscillator signals' pulses was developed.
- A concept for analyzing HOS using entropy features was presented. This concept uses FuzzyEn to assess the synchronization states of the oscillator circuit and to construct their entropy maps. The dependence of FuzzyEn on the embedding vector length *L* of SHR time series was studied to classify the synchronization states of the circuit.
- An algorithm for analyzing HOS of non-pulse signals was developed. Experimental data on local field potentials (LFP) of the rat hippocampus in a study of phase-phase coupling of θ- and γ-waves (Scheffer-Teixeira R and Tort A B L, 2016) were analyzed as an example.

The rest of the paper is organized as follows. Section 2 (Methods) presents the general workflow diagram of the method, describes the model circuit HOS (phase locking), and techniques for FuzzyEn. Section 3 (Results) is dedicated to numerical results for the pulse dynamics of the circuit without noise and under the impact of frequency-modulated noise. Section 3.4 analyzes the synchronization of signals with a limited spectrum using the example of LFP of the rat hippocampus. Sections 4 (Discussion) and 5 (Conclusion) summarize the proposed algorithm. The oscillator circuit calculations were performed in MatLab/Simulink, and the signal processing and synchronization analysis results were obtained using Python.

## 2 Methods

### 2.1 Workflow diagram of phase-locking entropy analysis

Figure 1 shows the general workflow diagram of the approach to entropy analysis of phase-locking dynamics. The initial step involves determining the type of signal used (*yellow square*): pulsed or non-pulsed. If the signals are pulsed (pulse1 and pulse2, *green squares*), they are multiplied (AND operation) to form phase-locking intervals. For spike signals (e.g., neuron spikes, ECG, etc.), pulse1 and pulse2 are formed by the sharp rise (or fall) of spike amplitudes. Then, within the phase-locking intervals, SHR values are formed from the ratios of the periods of both signals. The set of SHR values forms a time series of rational fractions $n/m$.

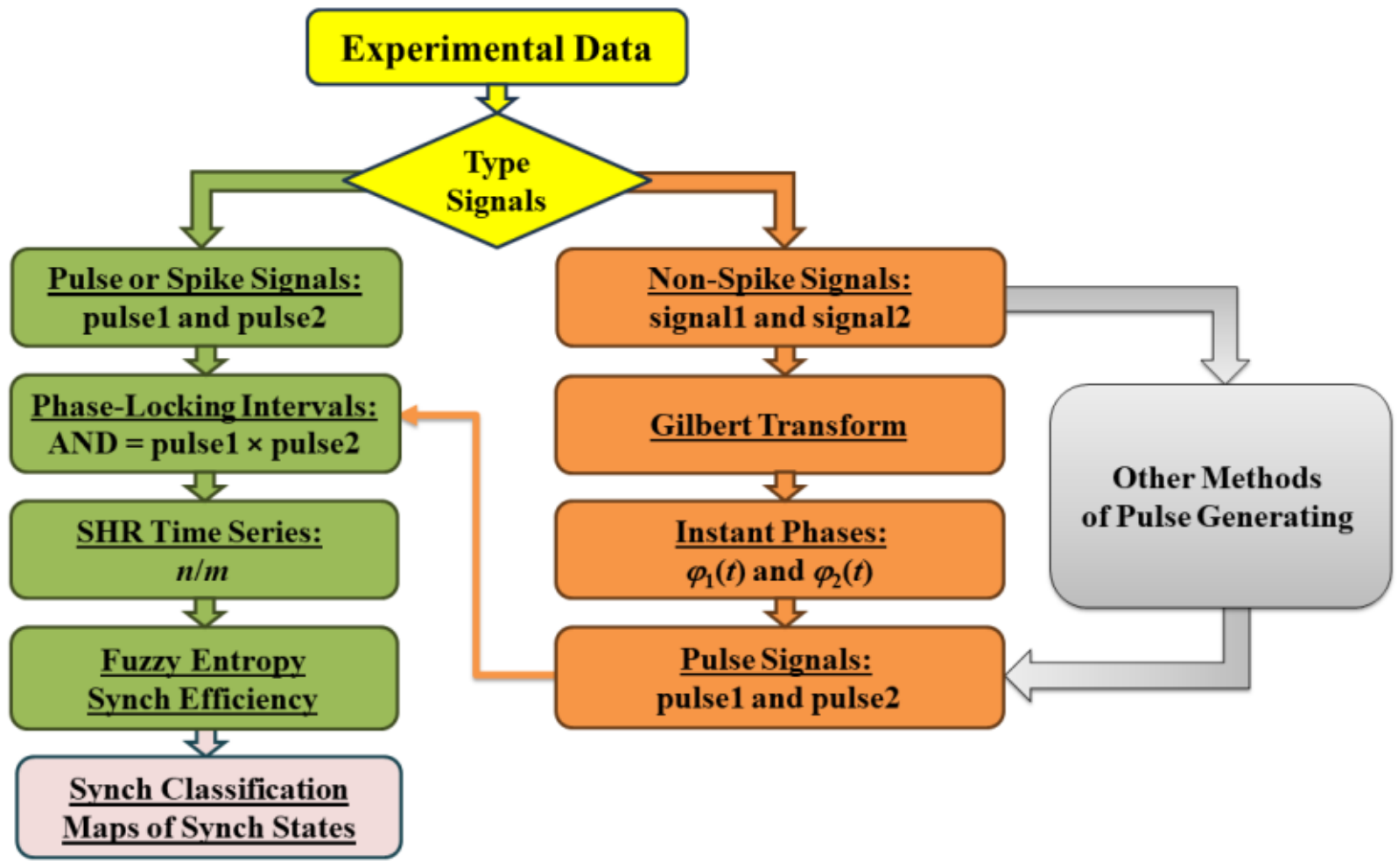

**Fig.1. Workflow diagram of phase-locking entropy analysis.**

For non-pulsed (non-spike) signals (*orange squares*), it is necessary to first construct phase functions $\phi_1(t)$ and $\phi_2(t)$ using the Hilbert transform, from which secondary pulse signals (pulse1 and pulse2) are formed. Alternative methods for forming pulse signals are shown on the diagram with *grey square* arrows. For the obtained SHR time series, FuzzyEn and SE are calculated. In the final step, the HOS states of the system are classified based on their values. A synchronous state corresponds to a low FuzzyEn and a high SE ~100%, while a non-synchronous state corresponds to a high FuzzyEn and a low SE <20%. Synchronization states are classified based on the dependence of FuzzyEn on the embedding vector length $L$ of the SHR time series.

## 2.2 Circuit of Resistively Coupled Pulse Oscillators

The object of synchronization modeling in our study is a digital-analog circuit of two resistively coupled pulse oscillators (Fig. 2a), with two resistors $R_{12}$ and $R_{21}$ placed between the inputs and outputs of the oscillators. Each oscillator is a simple self-oscillator based on two inverters with RC coupling (Fig. 2b). Due to positive feedback between the input of the first (node 1) and the output of the second inverter (node 2) in the circuit (Fig. 2b), rectangular pulses are generated, with periods and durations proportional to the time constants $T_1 = R_1 \cdot C$ and $T_2 = R_2 \cdot C$, respectively. In this circuit, the gate diodes $D_1$ and $D_2$ allow independent adjustment of the pulse periods and widths by changing the resistances $R_1$ and $R_2$. The dynamics of a single-resistor circuit without diodes are described in Appendix A (Boriskov P, 2024).

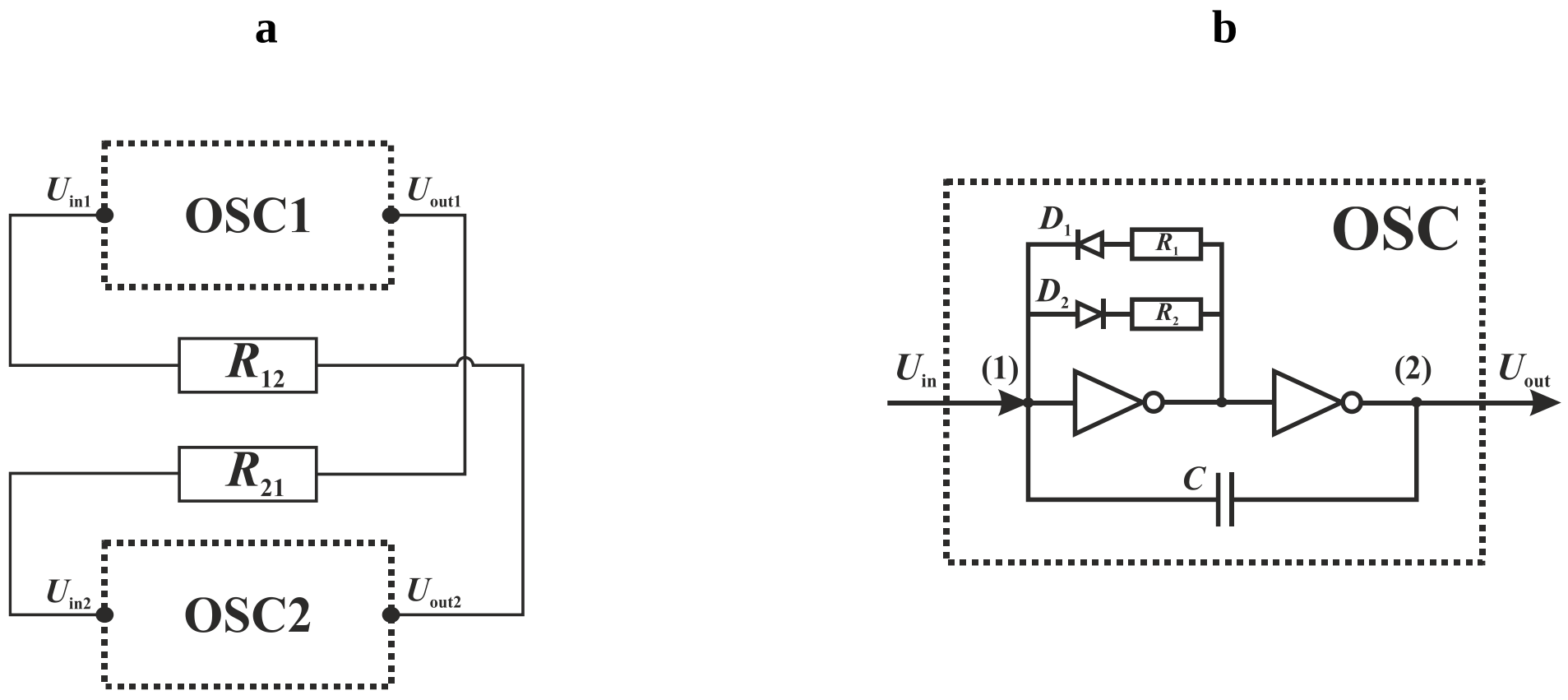

**Fig. 2 (a) The circuit of two resistively coupled pulse oscillators (OSC1 and OSC2). (b) The circuit of basic oscillator (OSC) with two NOT gates and RC coupling. The input (1) and output (2) nodes of the oscillator are indicated.**

**Table 1 Parameters of basic oscillator circuits**

| | OSC1 | OSC2 |
|---|---|---|
| $C$ nF | 100 | 100 |
| $R_1$ kΩ | 17 | 39 |
| $T_1$ ms | 1.7 | 3.9 |
| $R_2$ kΩ | 1 | 1 |
| $T_2$ ms | 0.1 | 0.1 |
| Logic NOT: $U_s$ = 5 V, $U_{low}$ = 1.5 V, $U_{high}$ = 3.5 V | | |

*Taken from Fig.2b.The NOT logic gate has a supply voltage ($U_s$) and low and high levels ($U_{low}$ and $U_{high}$)

The parameters of basic oscillator circuits (OSC1 and OSC2) used in all calculations are given in Table 1. Both oscillators have equal capacitances $C$ and resistances $R_2$, so the pulse widths are the same in the absence of resistive coupling. At the same time, the intrinsic pulse periods of the oscillators, determined by resistances $R_1$, differ by more than twice (~2.2, see Table 1). The oscillator with a shorter pulse period (higher frequency) is designated as the leading one (OSC1), while the oscillator with a lower frequency is the lagging one (OSC2).

Although a full analysis of the dynamics of the circuit shown in Fig. 2a is beyond the scope of this study, it is worth noting the key aspects important for subsequent analysis. The system of coupled oscillators generates output rectangular pulses OSC1 and OSC2 with controlled periods (frequencies) when changing resistances $R_{12}$ or $R_{21}$. The impact of one oscillator on the other manifests as additional currents at their inputs (node 1) through these resistances, reducing the frequency and thus slowing down the pulse rate. The lower the resistance $R_{12}$ or $R_{21}$, the greater the amplitude of the corresponding current $I_{12}$ or $I_{21}$, and the stronger the frequency suppression of the corresponding oscillator. The reverse effect of coupling currents on the oscillators through their outputs is practically absent, ensuring unidirectional current impact from the first oscillator to the second through $R_{12}$ and vice versa.

However, the pulse widths do not change significantly with changes in $R_{12}$ or $R_{21}$, which becomes more pronounced as the ratio $T_2/T_1$ decreases (no less than 17 times, see Table 1). Thus, changing the coupling between oscillators through resistances $R_{12}$ or $R_{21}$ leads not only to different temporal correlations (synchronization) of the output pulses but also controls their frequency modulation.

### 2.3 High-Order Synchronization Metrics of Pulse Oscillations

Phase synchronization is defined as the constancy of instantaneous phase relationships between signals, meaning no phase precession is present, and the phase of one signal always accurately predicts the phase of the other (Pikovsky Author A et al., 2002). The mathematical condition for phase locking $n$:$m$ of two signals with phases $\phi_1(t)$ and $\phi_2(t)$ is given by:

$$\Delta\varphi_{\mathrm{nm}} = \left|n\varphi_1(t) - m\varphi_2(t)\right| < \varepsilon \text{ for any } t\,, \quad (2)$$

where $n$ and $m$ are integers, and $\epsilon$ is a small number. Instantaneous phases of narrowband signals are determined based on analytical signals using the Hilbert transform (Rabiner L R et al., 1978). In general, including relaxation (pulse) signals, the wrapped phase $\phi(t)$ can be understood as a function proportional to the fraction of the period, increasing by $2\pi$ during one cycle of oscillations (switching):

$$\varphi(t) = 2\pi\frac{t - t_1}{t_2 - t_1},\ t_1 < t \le t_2, \quad (3)$$

where $t_1$ and $t_2$ are the start and end times of one cycle, which coincide with the neighboring switching times (either the ON→OFF or OFF→ON time) for pulses.

The effect of synchronization with different $n$ and $m$ is also called subharmonic synchronization (or HOS) because the spectra of both signals contain coinciding frequencies $m\omega_{01}=n\omega_{02}$, where $\omega_{01}$ and $\omega_{02}$ are the fundamental spectral frequencies of the first and second signals. The ratio $n$:$m$ is called the subharmonic ratio (SHR) (Velichko A et al., 2018). This ratio is evidently inversely proportional to the number of periods of each oscillator fitting into a certain interval. An example of HOS synchronization with SHR=2:5 satisfying condition (2) is shown later in the 'Results' section in Fig. 3.

HOS can be identified without evaluating condition (2). With an initial zero phase shift, one can simply multiply the pulses (Fig. 3a), i.e., assign them a pulse train that represents a logical multiplication function (pulse1 AND pulse2). If this train is a periodic sequence of single pulses, the original pulse signals are synchronized and satisfy condition (2). In quasi-regular synchronization of two signals, different SHR values occur in various phase-locking intervals, resulting in a time series of rational SHR values. Using the AND operation to assess the SHR of two pulse sequences, we determine the positions of their coincidences and then count the number of periods of the first signal (*n*) and the second signal (*m*) between them. Performing this procedure for each period of the multiplied pulse signal, we obtain a binary series consisting of pairs of n and m numbers. These pairs can also be assigned SHR values as rational fractions n/m. HOS will also be characterized by the synchronization efficiency metric (synch efficiency, SE), defined as the maximum probability $P_{\max}$ in the SHR distribution (Velichko A et al., 2018; Velichko A, 2019):

$$\mathrm{SE} = P_{\max}(n_1 : m_1) \cdot 100\ \%, \tag{4}$$

where $n_1$:$m_1$ is the ratio with the highest relative frequency in the SHR time series.

## 2.4 Fuzzy Entropy and Classification of Synchronization States

To assess the regularity of time series in this study, FuzzyEn was used, which is the most popular in various applied and theoretical entropy data analysis studies. The principle of FuzzyEn calculation is based on measuring the conditional probability that if vectors (segments of the time series) of length *L* are similar within an acceptable deviation, then vectors of length *L*+1 will also be similar within the same deviation. For a time series of length *N*, *N*−*L*+1 vectors can be compared. The similarity metric (correlation) between vectors is a fuzzy function that calculates the distances between a pair of compared vectors:

$$D_{ij}^{L} = \exp\left( -\frac{\left(d_{ij}^{L}\right)^{r_2}}{r_1} \right), \tag{5}$$

where $d_{ij}^{L} = \max\limits_{k\in(0,L-1)} \left| x_{i+k} - x_{j+k} \right|$ is the Chebyshev distance between vectors, defined as the maximum difference between its components. Using fuzzy similarity, the probability of similar patterns of length *L* appearing in the time series *N* is calculated:

$$\phi^{L}(r_1, r_2) = \frac{1}{N-L} \sum_{i=1}^{N-L} \left( \frac{1}{N-L-1} \sum_{j=1, j\neq i}^{N-L} D_{ij}^{L} \right), \tag{6}$$

The numerator is the sum of similarity degrees for each pair of vectors. The final value of fuzzy entropy is defined as the negative natural logarithm of the ratio of the sum of the similarity degrees of vectors of length *L* and *L*+1:

$$\mathrm{Fuzzy}(L, r_1, r_2) = \mathrm{Ln}\left( \frac{\phi^{L}(r_1, r_2)}{\phi^{L+1}(r_1, r_2)} \right), \tag{7}$$

where $r$ is the threshold for the similarity measure, which determines the similarity degree of the vectors. In this study, the parameter $r$ was chosen empirically, based on the specific application and characteristics of the time series being analyzed.

The main adjustable parameters of FuzzyEn (7) are the embedding vector length $L$, the inverse exponent $r_1$, and the exponent $r_2$. As standard parameters, the minimal dimension (length) of the vector $L = 1$ is most often used. The parameter $r_1$ defines the permissible deviation of elements and is taken within the range of $(0.005 \div 0.4)\cdot std$ , where std is the standard deviation of the series elements (by default $r_1 = 0.01\cdot std$ ). The power dependence exponent $r_2$ is by default equal to 1. In all calculations of this study, the default values of $r_1$ and $r_2$ are used.

To classify the synchronization states of the oscillator circuit, the FuzzyEn values of the SHR time series were analyzed. The analysis focused on the dependence of FuzzyEn on the embedding vector length $L$. Low FuzzyEn values indicate high synchronization, while high FuzzyEn values suggest low synchronization. By studying the FuzzyEn values for different embedding vector lengths, the synchronization states of the oscillator circuit were classified into different categories, providing a detailed understanding of the synchronization dynamics.

# 3 Results

## 3.1 Variants of HOS in Pulse Oscillator Circuit

Figures 3-5 show various dynamic modes of the coupled oscillator circuits observed with different resistance values $R_{12}$ and $R_{21}$ listed in Table 2. Figure 3 demonstrates complete HOS of pulses from two oscillators (Type 1). The oscillogram of the pulse multiplication (bottom of Figure 3a) is a strictly regular sequence of single pulses, between which there are 5 periods of OSC1 pulses and 2 periods of OSC2 pulses, i.e., SHR = 2:5. As shown in Figure 3b, the condition (2) for the modulus of the phase difference $\Delta\phi_{nm}$ with $m$=5 and $n$=2 calculated using (3) is satisfied for values of $\epsilon$ not exceeding 0.03 radians.

**Table 2 Parameters of coupling resistances of circuit Fig. 2a for three types of HOS.**

| Type synchronization | Figure | $R_{12}$ MΩ | $R_{21}$ MΩ |
|---|---|---|---|
| Regular HOS(Type 1) | Fig.3 | 1.1 | 0.4 |
| Regular chimeric HOS (Type 2: 2a and 2b) | Fig.4a | 1.1 | 1.7 |
| | Fig.4b | 1.1 | 1.6 |
| Quasi-regular chimeric HOS (Type 3) | Fig.5 | 1.4 | 0.9 |

*The oscillator parameters are in Table 1

a b

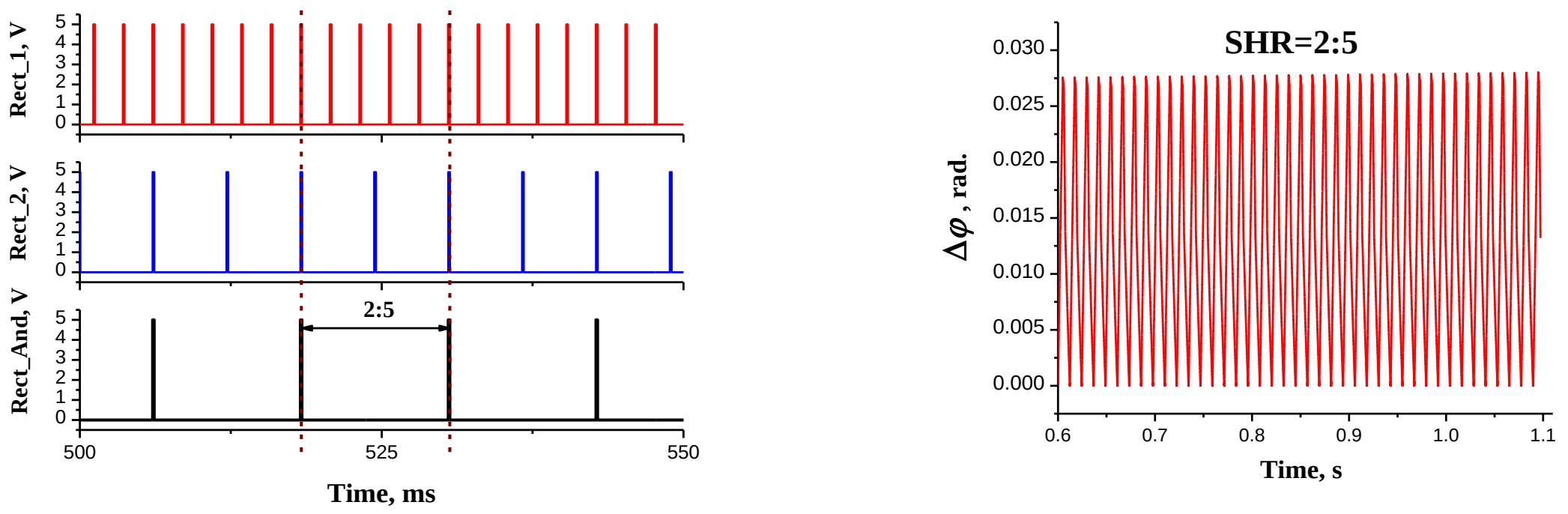

**Fig.3 Oscillograms of Type 1 HOS with SHR = 2:5 showing the output pulses of the master (Rect_1) and slave (Rect_2) oscillators and their multiplication (Rect_And). (b) The phase difference module Δ$\varphi$ with *m* = 5 and *n* = 2 of the output voltages Rect_1 and Rect_2. Calculation parameters are in Table 2.**

Figure 4 presents examples of the second type of HOS dynamics (Type 2) in two variants (2a and 2b), where the product of oscillator pulses forms periodic sequences grouped into batches of 5 (Figure 4a, Type 2a) and 16 (Figure 4b, Type 2b) pulses. In this case, the rows also represent periodic sets (with periods of 5 and 16) of SHR ratios 9:20, 4:9, 5:11, and 13:29, 9:20, 4:9 for variants 2a and 2b, respectively. For example, in variant 2a shown in Figure 4a, the SHR sequence has repeating sequences {9:20, 4:9, 5:11, 4:9, 9:20}.

a b

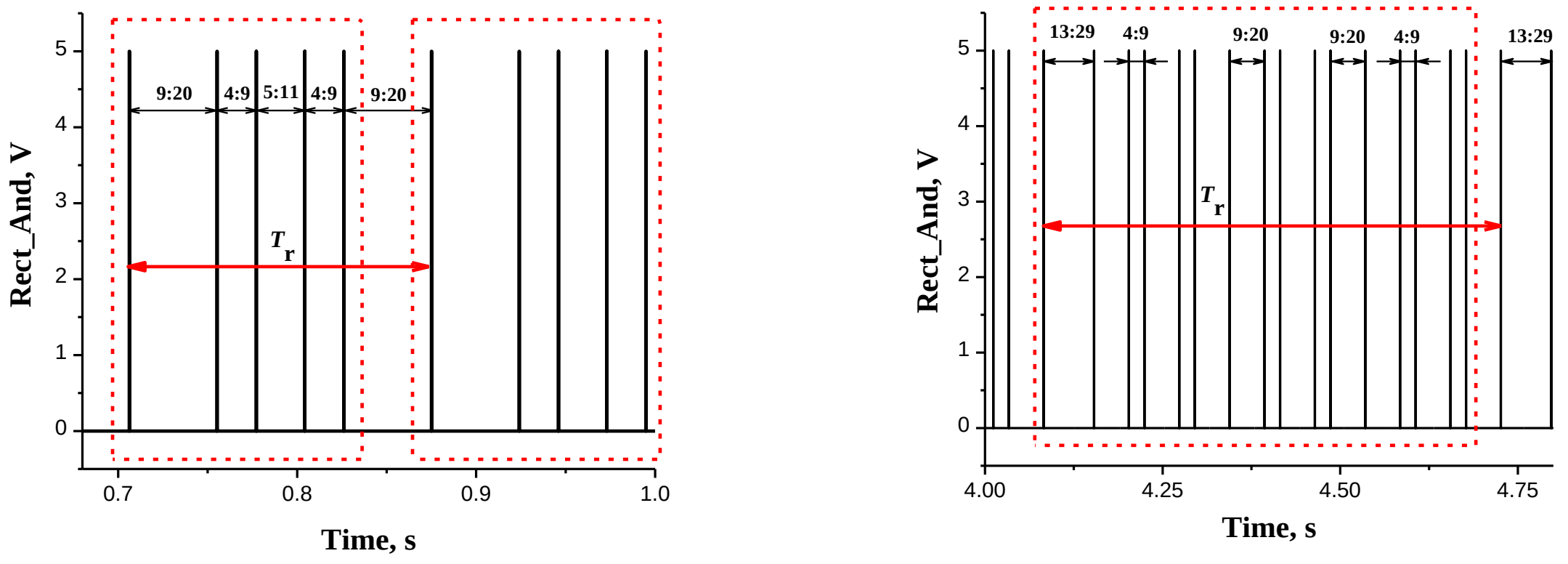

**Fig. 4. Oscillograms of multiplication of oscillators output pulses (Rect_And) for Type 2 HOS: (a) Type 2a and (b) Type 2b. Groups of pulses with periods $T_r$ (red arrows) are 5 (a) and 16 (b) indicated by red dashed rectangles with SHR sets (9:20, 4:9, 5:11) and (13:29, 9:20, 4:9), respectively. Calculation parameters are in Table 2.**

a

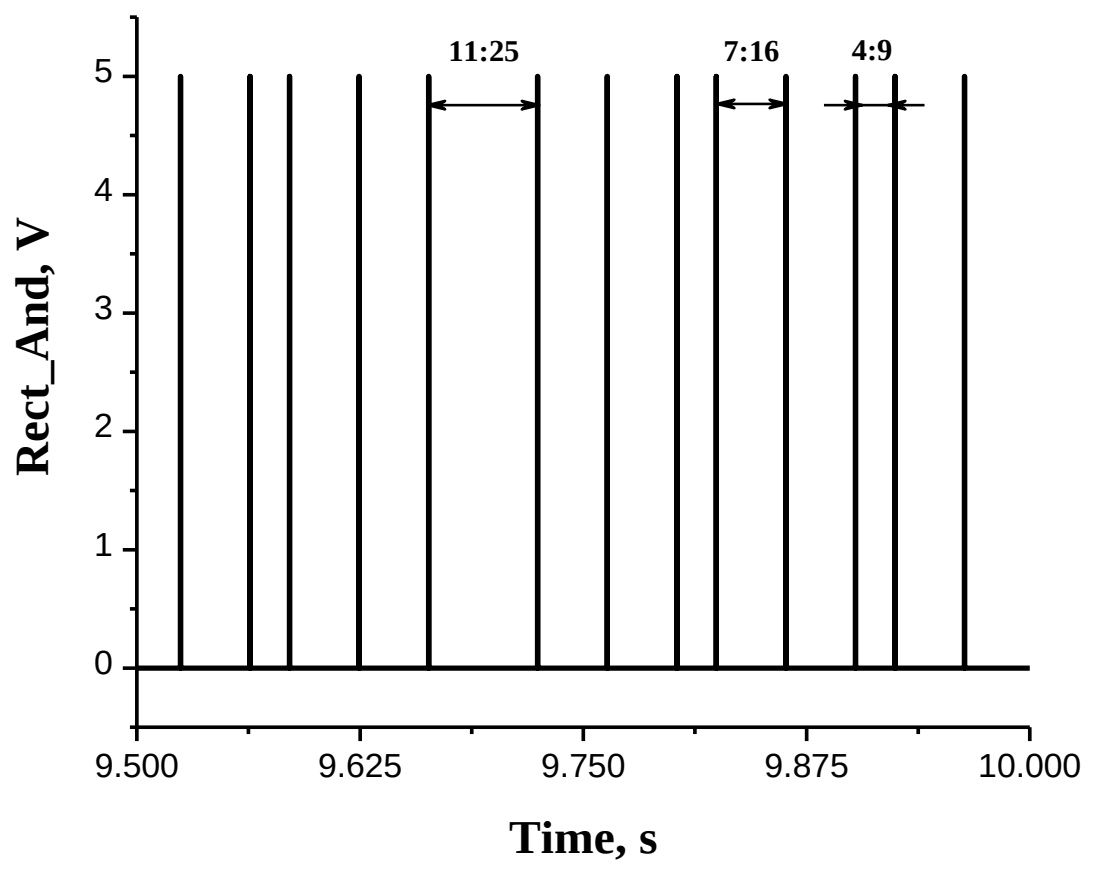

b

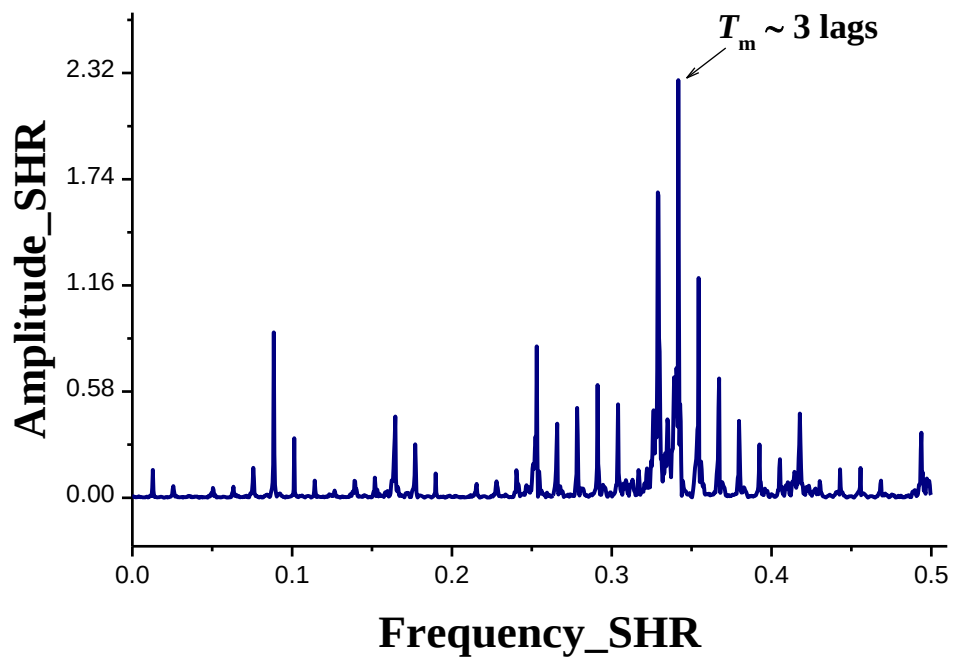

**Fig. 5. Oscillogram of multiplication of oscillators output pulses (Rect_And) (a) and AFR of SHR time series (b) for Type 3 HOS. The values from the set (11:25, 7:16, 4:9) of the SHR series and its period $T_m$ with the maximum amplitude of the spectrum are shown. Calculation parameters are in Table 2.**

The oscillograms in Fig.5a demonstrate the third type of circuit dynamics, where the output pulses of the oscillators form a quasi-periodic sequence of multiplication pulses. Unlike Type 2 HOS, this type has a quasi-periodic set of three SHR ratios: 11:25, 4:9, and 7:16. If we denote this set with the letters a, b, and c respectively, the sequence of SHR can be represented as {…cbccbccbc**a**cbccbc**a**cbccbc**a**cbccbccbccbc**a**cbccbc**a**…}. As seen, in HOS Type 3, the values repeat only within short intervals, and the regularity of the series is disrupted because SHR = 11:25 (letter a) is inserted in random positions. The amplitude-frequency response (AFR) of SHR (Fig.5b) shows the most frequent repetition interval $T_m$ lags, consisting of the sequence {7:16; 4:9; 7:16}, i.e., {cbc}.

We calculate the synchronization efficiency SE for three types of oscillator circuit dynamics. For complete HOS (type 1, Figure 3), it is obvious that SE = 100%. For Type 2, statistics give SE = 40% (Figure 4a) and SE = 56% (Figure 4b) with the most frequent value being SHR = 9:20. In the case of quasi-periodic synchronization of Type 3 (Figure 5), the most frequent value SHR = 7:16 has SE = 66%. If the first type of dynamics with SE = 100% represents true HOS according to condition (2), then the second and third types, where this condition is not met, can be called regular and quasi-regular chimeric HOS, respectively.

Thus, the dynamics of the coupled pulse oscillators (Figure 2) are classified into three types, among which the real HOS is the first variant (Figure 3). Only in this case does the strong coupling between oscillators ensure their synchronous interaction and fulfillment of condition (2). The dynamics of regular chimeric HOS Types 2a and 2b (Figure 4) are initiated by very weak interactions between the oscillators with periodic fluctuations of pulse periods and SHR series. Quasi-regular chimeric HOS (Type 3, Figure 5) results from weak interactions between oscillators, initiating fluctuations in pulse periods with disturbances in the periodicity of the SHR series. For Types 2 and 3 HOS, the phase difference of the pulses grows indefinitely for any $n$ and $m$, i.e., condition (2) is not met.

Fig.6 shows the dependence of FuzzyEn on the embedding vector length $L$ for the three types of HOS presented in Fig. 3-5. The complete HOS with SE = 100% (Fig.3) expectedly has zero entropy at any $L$ (black curve (1), Fig.6). In the case of the regular chimera HOS (Type 2), FuzzyEn is high at $L$ = 1 and only reaches zero as $L$ increases. The synchronization entropy of Type 2a (Fig.4a) is already zero at the second step of $L$ (Fig.6, red curve (2a)), whereas for Type 2b (Fig.4b), the entropy only drops to zero at $L$ = 11 (Fig.6, blue curve (2b)). This is evidently due to the different periodicity of the chimera synchronization: in the second variant, the SHR series has a significantly longer period and variation in SHR values. However, note that the periods (5 and 16) of these regular types of HOS do not match the values of $L$ at which FuzzyEn becomes zero (2 and 11, respectively).

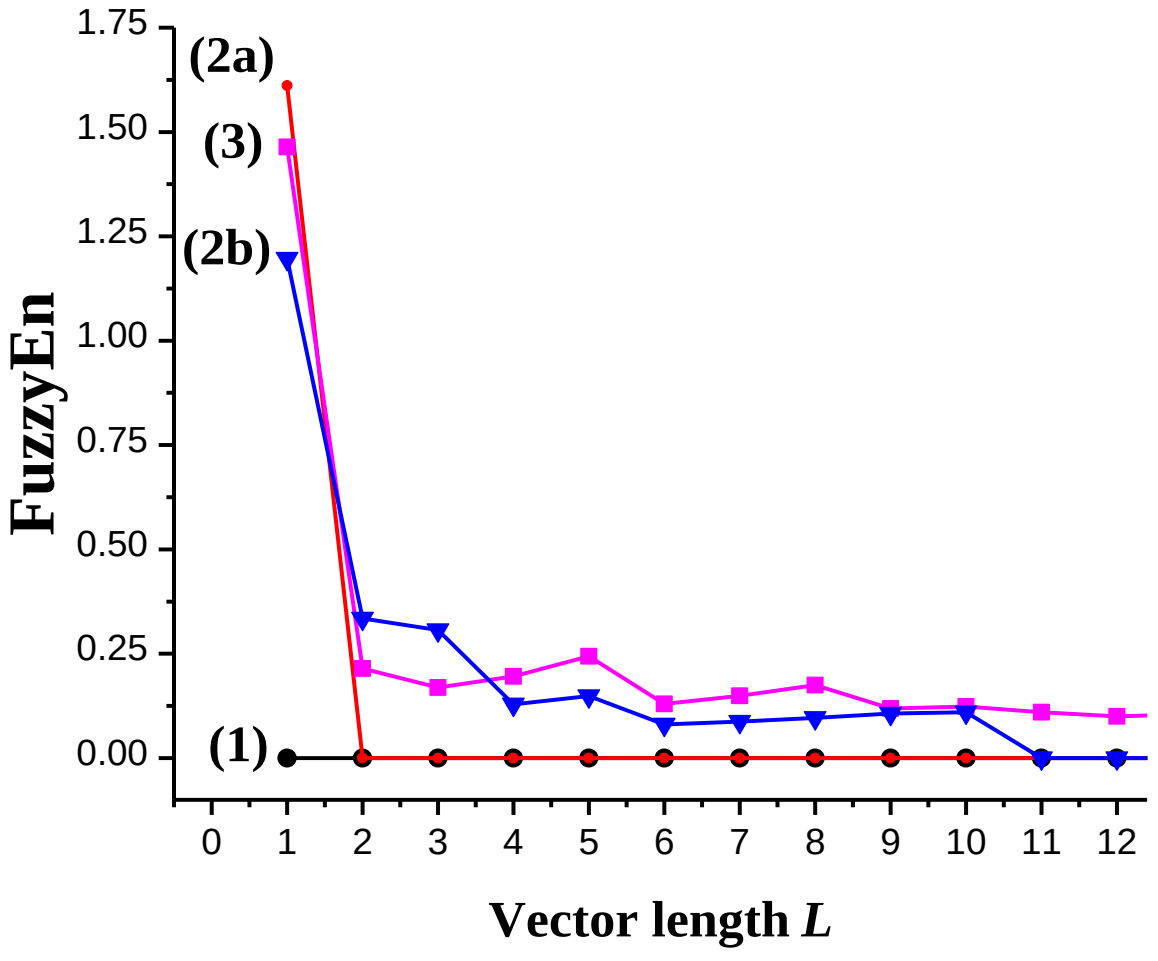

**Fig. 6. Dependencies of FuzzyEn on the embedding vector length *L* of SHR series for three types of HOS. Black curve (1) – Type 1, red curve (2a) and blue curve (2b) – Type 2, magenta curve (3) – Type 3.**

In the case of the quasi-periodic chimera HOS (Type 3, Fig.5), FuzzyEn does not drop to zero as the embedding vector length increases but instead reaches a constant value (Fig.6, magenta curve (3)). Thus, by varying *L*, the sensitivity of the FuzzyEn method to detecting synchronization states can be adjusted.

## 3.2 Distribution Maps of HOS States in Pulse Oscillator Circuit

In the oscillator circuit shown in Fig. 2a, for each pair of resistances $R_{12}$ and $R_{21}$ in the range from 0.2 to 2.1 MΩ with a step of 100 kΩ, SHR time series were calculated to assess the degree of synchronization.

### 3.2.1 Synchronization Efficiency Map

Figure 7 shows the map of SE distribution, indicating constant SHR values at 100% efficiency. In the left part of the map (small $R_{21}$ values), an area of HOS with a SHR ratio of 1:3 is observed. This area corresponds to the strong influence of the slow (follower) oscillator OSC2 on the fast (leader) oscillator OSC1 through a small $R_{21}$ value less than 0.3 MΩ.

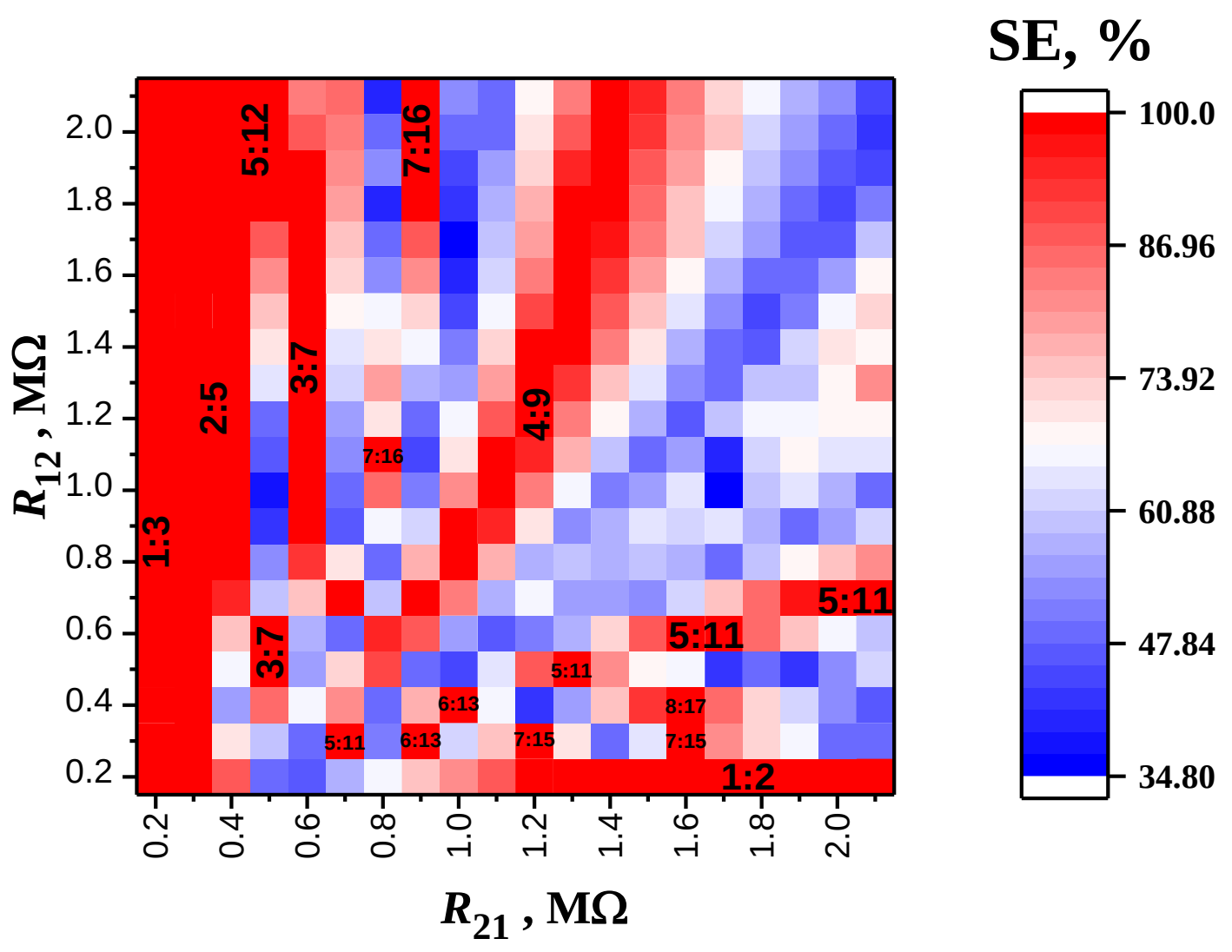

**Fig. 7. Distribution map of synchronization efficiency (SE). SHR values in areas with SE = 100% are indicated. The parameters of basic oscillators (Fig.2b) in Table 1.**

As the resistance $R_{21}$ increases, the influence of the follower oscillator decreases (the current $I_{21}$ drops), and areas of high HOS with SHR = 2:5 are observed on the map. Further to the right edge of the map, synchronization zones with other SHR values (3:7, 7:16, 4:9, etc.) are visible, where the SHR value gradually approaches the ratio of the oscillators' natural frequencies ($n/m \approx 0.4545$). The largest HOS area corresponding to SHR = 4:9 is located near the map's diagonal, where the frequency of the fast oscillator differs from the free oscillation frequency by less than 2%. Next to the SHR = 4:9 area, there is a diagonal zone with low SE, dividing the map into two parts, as shown by the blue and white-blue colors in Fig. 7. Below and to the left of this diagonal, there are small zones where the $n/m$ value exceeds 0.452, with SHR = 5:11, 6:13, 7:15, 8:17. In these areas, the follower oscillator's frequency begins to adjust to the leader oscillator's frequency, reducing the period of natural oscillations until the nearest SHR value of 1:2 is reached.

### 3.2.2 FuzzyEn Distribution Map of SHR Series

Figure 8 shows the FuzzyEn distribution maps of the SHR time series for embedding vector lengths $L$=1−4. On the map with $L$=1, it is noticeable that entropy exhibits a pronounced negative correlation with synchronization efficiency SE (see Fig. 7), i.e., in areas where SE approaches 100%, FuzzyEn approaches zero, and vice versa. The average Pearson correlation coefficient calculated from the SE and FuzzyEn distribution maps with $L$=1 is -0.903, indicating a high negative correlation between entropy and synchronization efficiency.

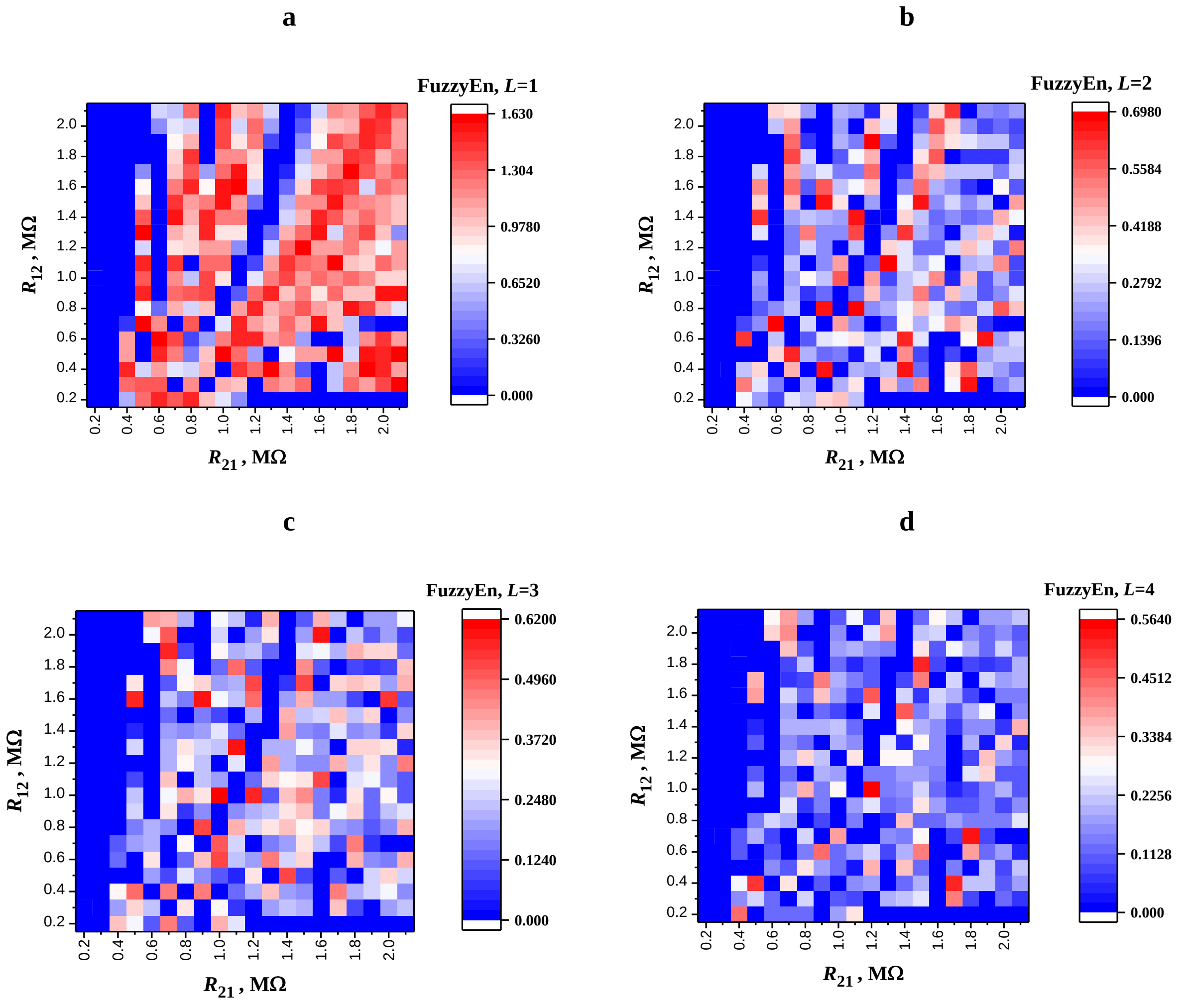

**Fig. 8. Distribution maps of FuzzyEn for embedding vector lengths of SHR series: *L*=1 (a), *L*=2 (b), *L*=3 (c), and *L*=4 (d). The parameters of basic oscillators (Fig.2b) are given in Table 1.**

There are clearly visible blue areas on the maps (Fig. 8) where FuzzyEn is close to zero, consistent with SE = 100% (red area, Fig. 7). As the vector length $L$ increases from 2 to 4, the maps show a decrease in entropy and an increase in blue areas, so that at $L$ = 4 only isolated red areas of high FuzzyEn remain (Fig. 8d). It could be stated that increasing the length of the embedding vector $L$ allows highlighting on the maps the most unsynchronized areas of dynamics and obtain the additional information about the areas of partial (quasi-periodic) HOS of oscillators.

### 3.3 HOS in the Pulse Oscillator Circuit Under Noise Influence

Let's consider the effect of HOS in the presence of white noise, which exclusively affects the frequency modulation of the oscillators, initiating fluctuations in pulse periods relative to their average values. We include two independent white noise generators in the circuit (Fig. 2a), each randomly varying the resistance $R_1$ of its oscillator (Fig. 2b) and the time constants $T_1=R_1 \cdot C$, which determine the pulse periods. The Simulink circuit Fig. 2a of each basic oscillator (OSC1 and OSC2) includes a variable resistance module, previously developed by us (see Appendix B (Boriskov P, 2024)), controlled by a noise signal $U_n(t)$ according to the linear law:

$$R_1(t) = R_o + K_n \cdot U_n(t)\,, \tag{8}$$

where $R_o$ is the reference resistance, and $K_n$ is the white noise gain coefficient. The discrete white noise $U_n(t)$ had standard parameters with a mean of zero and power $P_n$ determined from the variance $D_n$ and sampling step $t_n$:

$$P_n = \sqrt{D_n \cdot t_n}\,, \tag{9}$$

The noise sampling step in the simulation was chosen as $t_n$ = 0.01s, which is approximately half the average period of the master oscillator's pulses. The case of identical noise generators affecting the basic oscillators is only analyzed. The parameters of the noise generators ($K_n$ and $P_n$) are in Table 3.

Figure 9 shows the distribution maps of SE and FuzzyEn for an embedding vector length $L$=1 under white noise influence. As expected, these maps are also inverse to each other but become blurred compared to the synchronization state maps without noise (compare with Fig. 7 and 8a).

**Table 3: Parameters of oscillator circuit Fig.2a under noise with power (9).**

| | | | | |
|---|---|---|---|---|
| **Fig.9:** $K_n$ = 10 kΩ/W and $P_n$ = 24 mW | | | | |
| **Fig.10:** $R_{12}$ = 1.2 MΩ, $R_{21}$ = 0.2 MΩ, $K_n$ = 10 kΩ/W | | | | |
| **Fig.10b, curves (1)-(5):** | | | | |
| **(1)** $P_n$ = 10 mW | **(2)** $P_n$ = 14 mW | **(3)** $P_n$ = 24 mW | **(4)** $P_n$ = 27 mW | **(5)** $P_n$ = 45 mW |

* The parameters of basic oscillator circuits (Fig.2b) are in Table 1, except for the resistances ($R_1$(t)), which are controlled according to formula (8). The reference resistance $R_o$ (8) is equal $R_1$ of Table 1 for OSC1 and OSC2 oscillators, respectively

High synchronized states of the oscillator circuit under noise are maintained only at the left edges of the maps and have maximum SE values around 80% (red color in Fig. 9a) and minimum FuzzyEn values around 0.62 (blue color in Fig. 9b). These values approximately correspond to the average SE (white-red color in Fig. 7) and FuzzyEn (white-blue color in Fig. 8a) in the oscillator circuit without noise.

An example of HOS fluctuations in the SHR series under noise power $P_n$=24mW is shown in Fig. 10a. The level of SHR=1:3 is clearly visible, which corresponds to full HOS in the absence of noise. The dependencies of FuzzyEn on the embedding vector length $L$ for different noise power levels are shown in Fig. 10b. For the smallest noise powers (curves (1) and (2)), FuzzyEn reaches a steady low level for $L$>1. As the noise power increases, entropy increases and begins to oscillate weakly around the mean level, which does not decrease with increasing $L$ (curves (3) and (4)). At the highest noise levels, FuzzyEn oscillations become unstable, with a sharp increase when $L$ reaches larger values (curve (5), $L$=10).

a　　　　　　　　　　b

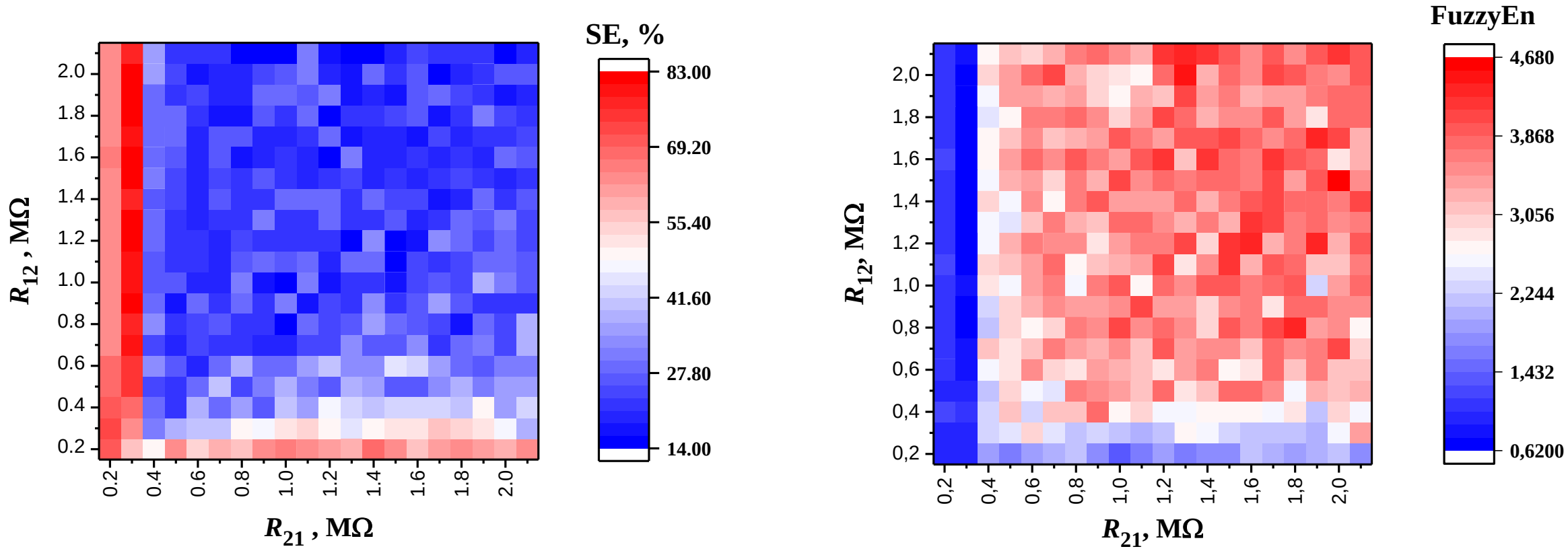

**Fig. 9. Distribution maps of SE (a) and FuzzyEn (b) under noise influence. Calculation parameters are in Table 3.**

a　　　　　　　　　　b

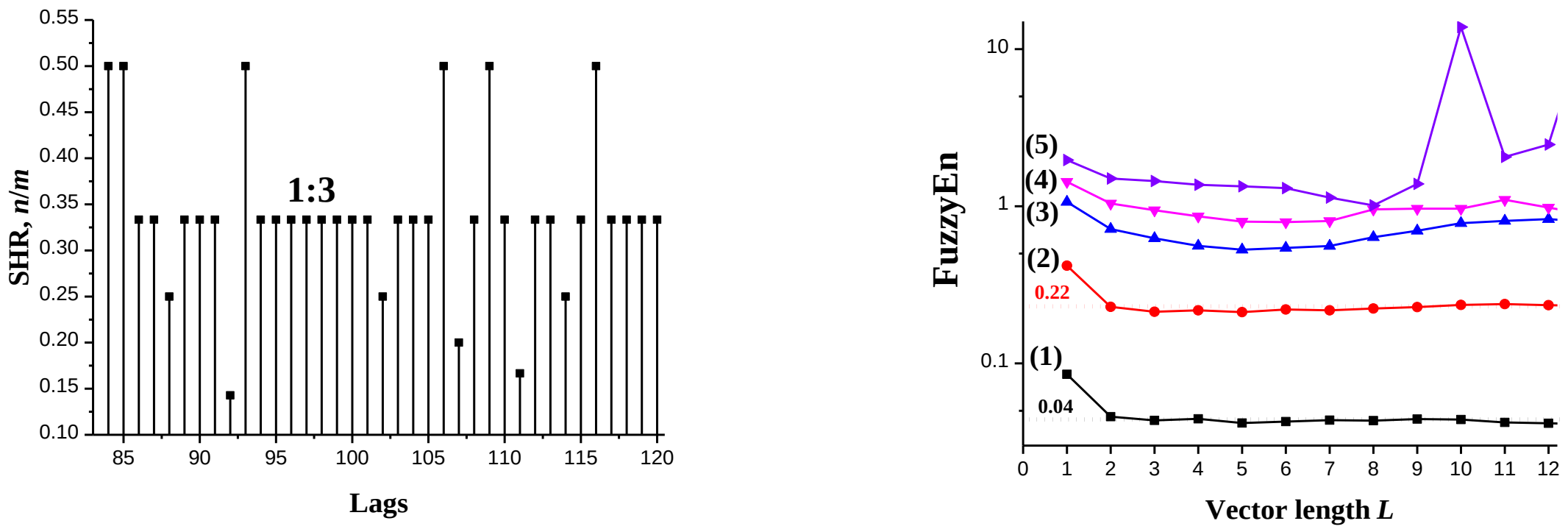

**Fig.10. (a) SHR time series (ratios n/m) under noise power $P_n$=24mW. SHR values at the level of 0.3(3) (1:3) are marked. (b) Dependencies of FuzzyEn (logarithmic scale) on the embedding vector length $L$ for increasing noise power: curves (1)-(5). Stationary FuzzyEn levels for curves (1) and (2) are shown. Calculation parameters are in Table 3.**

Noise affects oscillators' dynamics evidently lead to HOS failures and a gradual increase in FuzzyEn with increasing noise power. For the smallest noise powers (curves (1) and (2), Fig. 10b), the FuzzyEn dependencies on *L* resemble curves (2a) and (2b) in Fig. 6, describing type 2 HOS. As noise power increases, HOS failures increase, and the regularity of SHR series decreases further. In this case, with increasing *L*, entropy begins to oscillate slightly but does not fall to a low level (curves (3) and (4), Fig. 10b), resembling quasi-regular HOS Type 3 (blue curve (3), Fig. 6). The highest noise power initiates the destruction of the oscillators' HOS. Against the background of large fluctuations in the SHR series, entropy calculation instability arises (curves (5)-(7)): with increasing *L*, the amplitude of FuzzyEn oscillations begins to increase (curve (5), Fig. 10b). Such a self-excitation mode is characteristic of FuzzyEn for any highly irregular (chaotic) signal. Thus, entropy instability with increasing embedding vector length indicates strong chaos in the SHR time series and loss of synchronization in the circuit under sufficiently high noise power.

### 3.4 Entropy Analysis of HOS for Non-Spike Signals

So far, our studies have analyzed pulse-type output signals generated by the model circuit in Fig. 2. We extend the entropy analysis method to non-pulse (non-spike) signals with limited spectral width. We take two smooth, differentiable signals and transform them into analytical signals using the Hilbert transform (Rabiner L R et al., 1978). From the analytical signals, we extract phase functions of the wrapped type, i.e., instantaneous phases $\phi_1(t)$ and $\phi_2(t)$, representing rotation angles in the range $[-\pi, \pi]$. To the phase switching moments from $\pi$ to $-\pi$, we assign rectangular pulses of short duration relative to the average pulse periods (see Fig. 1). Thus, for the initial non-pulse signals, two pulse trains will be formed based on the phases $\phi_1(t)$ and $\phi_2(t)$. Next, the SHR series is calculated for these pulse trains using the AND operation, and the synchronization indicators SE and FuzzyEn are evaluated.

As an example, consider the experimental LFP data of the rat hippocampus from the database (Scheffer-Teixeira R and Tort A B L, 2016). It is known that LFP brain activity signals include several rhythms: $\theta$ (4-8 Hz), $\alpha$ (8-14 Hz), $\beta$ (14-30 Hz), and $\gamma$ (30-150 Hz). The gamma rhythm is further divided into three subranges: $\gamma_s$, $\gamma_m$, and $\gamma_f$, with the most significant being the $\gamma_s$-rhythm in the 30-50 Hz range. There is a hypothesis about phase-phase coupling between $\theta$- and $\gamma_s$-rhythms of the hippocampus, indicating synchronized activity of individual hippocampal neurons. This hypothesis is mostly confirmed by studies (Belluscio M et al., 2012; Xu X, Zheng C and Zhang T, 2013; Zheng C et al., 2016), although there are arguments against it, primarily of a technical nature (Scheffer-Teixeira R and Tort A B L, 2016). Extract the θ- and $\gamma_s$-oscillations from the LFP signal by passing it through a digital Butterworth band-pass filter in the θ- and $\gamma_s$-rhythm ranges. The advantage of the Butterworth filter is its uniform amplitude-frequency response (AFR) within the passband. To ensure a sharp roll-off at the band edges, the highest possible filter order should be used (up to the self-oscillation mode), which is 4 and 7 for the θ- and $\gamma_s$-ranges, respectively. Additionally, we will apply the well-known method of double (forward and backward) passage of the LFP signal through the filter to suppress nonlinear phase distortions.

Apply the Hilbert transform to the obtained $\theta$- and $\gamma_s$-oscillations to form phase functions ($\theta$- and $\gamma_s$-phases) and pulse trains ($\theta$- and $\gamma_s$-pulses) using the algorithm described above, demonstrated by the example in Fig. 11. Finally, for the $\theta$- and $\gamma_s$-pulses, calculate the SHR time series and evaluate FuzzyEn at different embedding vector lengths and synchronization efficiency SHR.

Figure 12a shows the dependencies of FuzzyEn on the embedding vector length $L$ for the LFP $\theta$- and $\gamma_s$-signals (from six rats) randomly selected from the database (Scheffer-Teixeira R and Tort A B L, 2016). As seen, the entropy curves split into two groups. In the first group (curves (1)-(3)), entropy decreases with increasing $L$ practically to zero, indicating an analogy with HOS Type 2 (compare with curve (2b) of Fig. 6) and some connection between the $\theta$- and $\gamma_s$-rhythms. The second group of FuzzyEn curves (4)-(6) indicates the presence of strong noise in the LFP signals, as entropy has a large oscillation amplitude and remains at a high level. LFP noise may be due to technical reasons (LFP registration noise, etc.) or other factors, such as the brain activity of animals (fluctuations in excitation and rest intervals). Since the first group revealed rhythm synchronization, in our opinion, the entropy curves (4)-(6) are simply unsuccessful registrations of LFP signals with a strong noise background. This is also proven by the synchronization efficiency assessments shown in Fig. 12b. The maximum relative frequencies in the distributions are found for SHR = 1:7 in all six data sets, but SE values are significantly lower for heavily noise-contaminated LFP signals (curves (4)-(6), Fig. 12a).

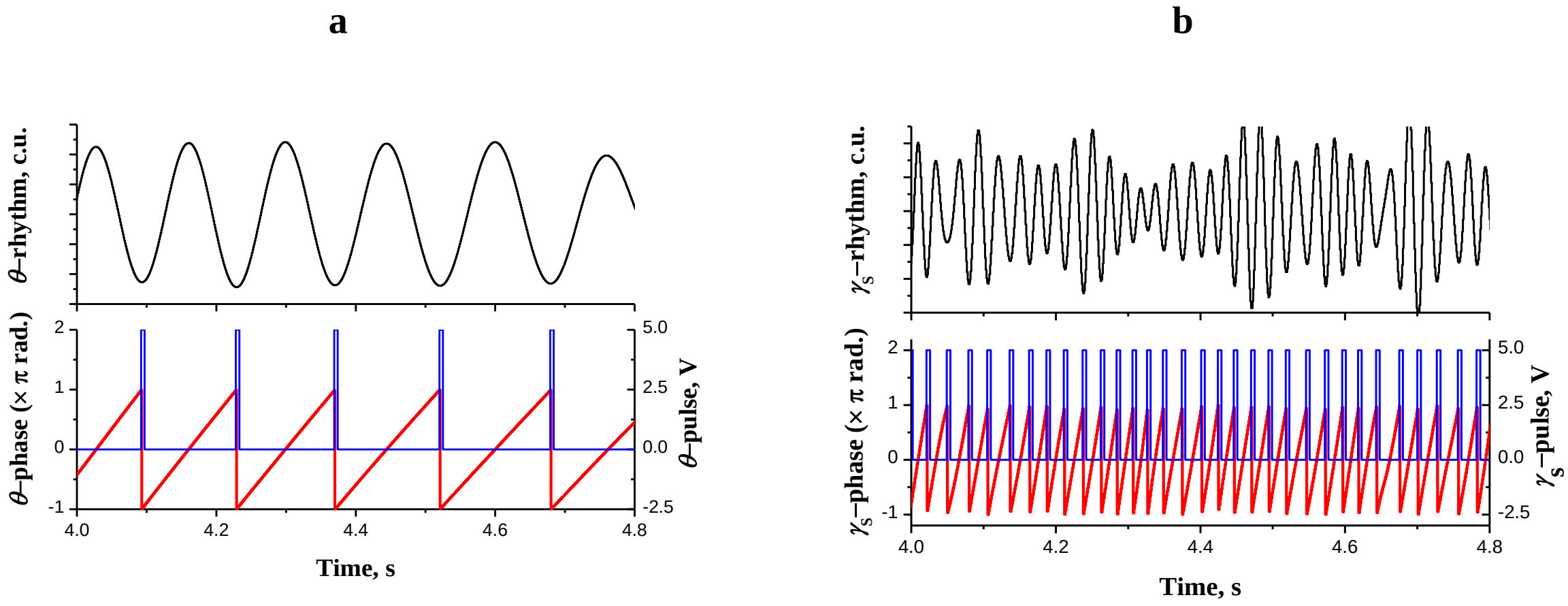

**Fig.11. LFP oscillograms (black curves, conventional units), phases (red curves) and pulse trains (blue curves) of θ-rhythm (a) and $\gamma_s$-rhythm (b). Experimental LFP of the rat hippocampus is from the work (Scheffer-Teixeira R and Tort A B L, 2016).**

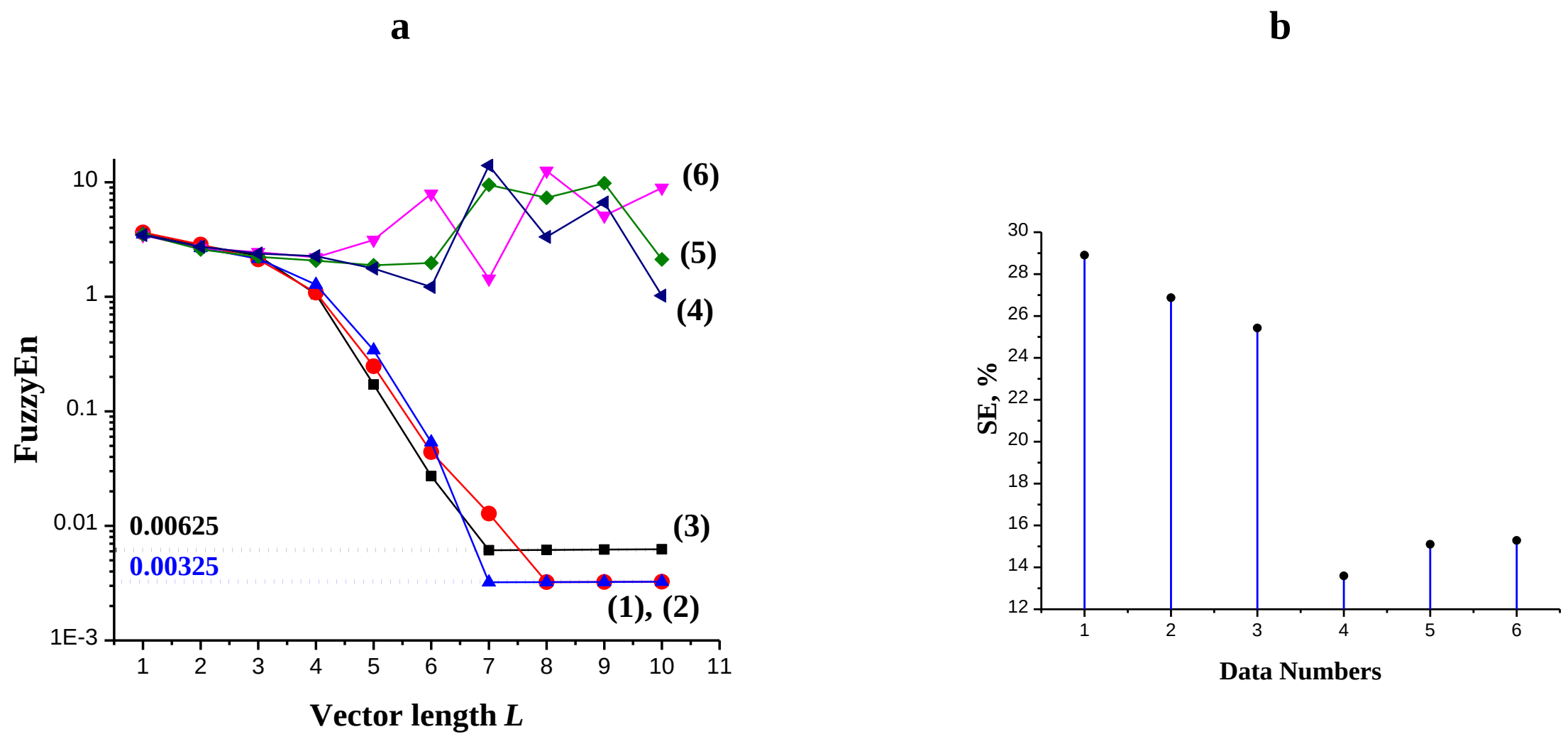

**Fig.12. (a) Dependencies of FuzzyEn on the embedding vector length *L* of SHR time series for θ and $\gamma_s$ LFP signals. (b) Distributions of SHR values for θ- and $\gamma_s$-pulses. Calculation parameters are in Table 3.**

This tendency for FuzzyEn to split into these two groups is also observed in other LFP datasets from the database (Scheffer-Teixeira R and Tort A B L, 2016), with a predominance of entropy dependencies from the second group, indicating a strong noise component in the LFP signals. Therefore, the analysis of HOS in LFP suggests a possible weak connection between $\theta$- and $\gamma_s$-rhythms, indicating some degree of rhythm synchronization. However, since brain activity and the network system inherently operate within a stochastic framework, LFP data require more precise recording and processing, including statistical analysis, and, overall, a larger number of recordings and sample sizes.

## 4 Discussion

Let's summarize the proposed method presented in the diagram in Figure 1. Depending on the signal type, pulse trains are formed from experimental data, which are used to analyze the nature and degree of HOS. If the signals are pulse or spiking (green squares in Figure 1), pulses of constant duration are generated at the moments of switching or sharp transitions, such as from low to high values. The formation of pulse trains for non-pulse signals (with limited spectral width) is based on the instantaneous phases obtained using the Hilbert transform, as shown in the diagram in Figure 1 (orange squares). In this case, pulses in the sequences are generated at the moments of phase switching from $\pi$ to $-\pi$.

Alternative methods for forming pulse trains from non-pulse signals (Figure 1, gray square) can also be considered. For example, a simple method is generating pulses at the moments of maximum values (peaks) of oscillating signals. However, it is necessary to monitor false peaks initiated by noise, which can significantly distort the phase-locking dynamics of the studied system.

Next, using the AND procedure, SHR time series in the form of rational fractions of pulse periods *n*/*m* are formed, and SE and FuzzyEn are calculated. The ultimate goal of entropy analysis can be a simple classification of phase locking *n*:*m*, i.e., determining whether HOS exists or not, which is decided by assessing SE or FuzzyEn with *L*=1, for example, by some threshold value.

Using SE and FuzzyEn, it is also possible to create maps of HOS states of the analyzed systems controlled by a set of parameters. A more complex assessment of the degree and nature of synchronization is carried out with increasing lengths of embedding vectors of the SHR series. For example, by increasing the vector length, it is possible to highlight the most unsynchronized areas of the system dynamics on FuzzyEn maps.

Let's consider the structure of the SHR time series, on the basis of which entropy is calculated in this approach, and conclusions about the synchronization of the system are drawn. In the case of partial (chimeric) HOS (Figures 4 and 5) or under the influence of moderate noise (curves (2)-(4), Figure 10b), the ratios of oscillator periods fluctuate around average values with small dispersion. For example, the relative standard deviation of the *n*/*m* ratios in the SHR series shown in Figure 4a does not exceed 1.2%. Meanwhile, these three pairs of periods from the set {9:20; 4:9; 5:11} differ significantly from each other. Also, in Figure 4, a strong difference in periods can be seen in the oscillograms of the multiplication (Rect_AND) of the output pulses of the oscillators. It could be said that even small fluctuations in the frequency of oscillators lead to a strong spread of distances between the moments of pulse coincidences. As a result, there is a strong difference in integer SHR pairs in different time intervals, i.e., in the periods of the multiplication signal.

In our opinion, SHR time series most adequately reflect the dynamics of synchronization. A time series composed of pulse periods of the multiplication (AND) signal could also be used for entropy analysis of HOS. The disadvantage of this method is that the ratio of periods strongly depends on the frequency (period) fluctuations of pulse signals, and even under weak noise impact, it will have high dispersion and high entropy. In this case, the synchronization pattern will be heavily noisy, complicating its analysis.

Using the ratio of the number of periods instead of the period values in the SHR series is analogous to filtering, where frequency fluctuations are averaged. A more detailed analysis of this issue can be carried out in future studies.

The resistively coupled oscillator circuit demonstrates three phase-locking modes (Table 1): full regular HOS (Type 1), regular chimeric HOS (Type 2), and quasi-regular chimeric HOS (Type 3). Apparently, this circuit does not exhibit another known type of dynamics — deterministic chaos, which would manifest in HOS as a completely chaotic SHR series. The entropy characteristic of such a mode should be similar to the dynamics of HOS with high noise levels (see Fig. 10b, curve (5)). FuzzyEn should be high and quickly enter a self-excitation mode as $L$ increases.

The self-excitation mode of FuzzyEn, where oscillations with increasing amplitude appear with increasing embedding vector length, is due to the loss of algorithm stability (equations (5)-(7)) for finite irregular time series. With a sufficiently long embedding vector $L$, no similar patterns of length $L$+1 may be encountered. However, since fuzzy similarity (equation (5)) is used in the calculation of FuzzyEn, the probability for patterns of length $L$+1 may be very small but not zero. As a result, the ratio of probabilities for patterns of length $L$ and $L$+1 can suddenly increase for certain values, leading to a rise in FuzzyEn. Thus, the self-excitation of FuzzyEn is caused by the loss of statistical stability of the calculation as $L$ increases for a highly noisy (chaotic) finite series. This instability can only be overcome by increasing the length $N$ of the series itself to a size where the probabilities of finding similar long patterns become statistically significant. For a regular (or nearly regular) finite series, the algorithm's self-excitation cannot occur as long as $L<N$ and FuzzyEn approaches zero as $L$ increases, because the probabilities of finding similar patterns of length $L$ and $L$+1 become identical.

The conducted study shows that the nature of the dependence of FuzzyEn on the embedding vector length $L$ can provide valuable information about the synchronization dynamics in the system. The loss of FuzzyEn stability with increasing $L$ may indicate the presence of chaotic dynamics in the SHR series and a lack of synchronization. Conversely, stable FuzzyEn behavior indicates regularity and predictability of the time series dynamics, with the system being in a synchronized state. Additionally, a decrease in FuzzyEn with increasing $L$ may suggest that the series contains hidden regularity, which becomes more pronounced at higher dimensions of the embedding vector.

## 5 Conclusions

The proposed approach to assessing HOS signals can be useful in various fields of science and technology, such as evaluating and analyzing the synchronization of financial, meteorological, and physiological data, including biomedical diagnostics. Its main advantage over other methods is the simplicity of the obtained characteristics in the form of fractional-rational SHR sequences, which can be obtained interactively. Additionally, the digital signals (rectangular waves) used in the method make it promising for implementation in fully digital platforms, such as FPGA-based systems. Further research should aim to enhance the sensitivity of the method to minor changes in synchronized signal dynamics, potentially using other entropy indicators and machine learning algorithms.

## References

Alter O, Brown P and Botstein D 2000 Singular Value Decomposition for Genome-Wide Expression Data Processing and Modeling *Proc Natl Acad Sci U S A* **97** 10101–6 https://doi.org/10.1073/pnas.97.18.10101

Becker D 2006 Fundamentals of Electrocardiography Interpretation *Anesth Prog* **53** 53–63; quiz 64

https://doi.org/10.2344/0003-3006(2006)53[53:FOEI]2.0.CO;2

Belluscio M, Mizuseki K, Schmidt R, et al., 2012 Cross-Frequency Phase-Phase Coupling between Theta and Gamma Oscillations in the Hippocampus *J Neurosci* **32** 423–35 https://doi.org/10.1523/JNEUROSCI.4122-11.2012

Biagi P, Piccolo R, Ermini A, et al.,2001 Possible earthquake precursors revealed by LF radio signals *Natural Hazards and Earth System Science* **1** https://doi.org/10.5194/nhess-1-99-2001

Bonet-Jara J, Quijano-Lopez A, Morinigo-Sotelo D, et al., 2021 Sensorless Speed Estimation for the Diagnosis of Induction Motors via MCSA. Review and Commercial Devices Analysis *Sensors* **21** https://doi.org/10.3390/s21155037

Boriskov P 2024 Chaotic discrete map of pulse oscillator dynamics with threshold nonlinear rate coding *Nonlinear Dyn* **112** 1–17 https://doi.org/10.1007/s11071-023-09217-9

Chen W, Wang Z, Xie H, et al., 2007 Characterization of surface EMG signal based on fuzzy entropy *IEEE Trans Neural Syst Rehabil Eng* **15** 266–72 https://doi.org/

Chou J, Bramhavar S, Ghosh S, et al., 2019 Analog Coupled Oscillator Based Weighted Ising Machine *Sci Rep* **9** https://www.nature.com/articles/s41598-019-49699-5

Delgado-Bonal A and Marshak A 2019 Approximate entropy and sample entropy: A comprehensive tutorial *Entropy* **21** 541 https://doi.org/10.3390/e21060541

Ishikawa A and Mieno H 1979 The fuzzy entropy concept and its application *Fuzzy Sets Syst* **2** 113–23

https://doi.org/10.1016/0165-0114(79)90020-4

Jiang H, Chen J, Dong G, et al., 2015 Study on Hankel matrix-based SVD and its application in rolling element bearing fault diagnosis *Mech Syst Signal Process* **52** https://doi.org/10.1016/j.ymssp.2014.07.019

Lowet E, Roberts M, Bonizzi P, et al., 2016 Quantifying Neural Oscillatory Synchronization: A Comparison between Spectral Coherence and Phase-Locking Value Approaches *PLoS One* **11** e0146443 https://doi.org/10.1371/journal.pone.0146443

Mallick A, Bashar M K, Truesdell D, et al., 2020 Using synchronized oscillators to compute the maximum independent set *Nat Commun* **11** 4689 https://doi.org/10.1038/s41467-020-18445-1

Nikonov D E, Csaba G, Porod W, et al., 2015 Coupled-Oscillator Associative Memory Array Operation for Pattern Recognition *IEEE Journal on Exploratory Solid-State Computational Devices and Circuits* **1** 85–93 https://doi.org/10.1109/JXCDC.2015.2504049

Park J, Mackay S and Wright E 2003 Preface *Practical Data Communications for Instrumentation and Control* (Oxford: Newnes) pp xi–xiii https://doi.org/10.1371/journal.pone. 0146443

Pikovsky Author A, Rosenblum M, Kurths J, et al., 2002 Synchronization: A Universal Concept in Nonlinear Science *Am J Phys* **70** 655 https://doi.org/10.1523/ JNEUROSCI.4122-11.2012

Rabiner L R, Gold B and Yuen C K 1978 Theory and Application of Digital Signal Processing *IEEE Trans Syst Man Cybern* **8** 146 https://doi.org/ 10.1109/TSMC.1978.4309918

Romera M, Talatchian P, Tsunegi S, et al., 2018 Vowel recognition with four coupled spin-torque nano-oscillators *Nature* **563** https://doi.org/10.1038/s41586-018-0632-y

Scheffer-Teixeira R and Tort A B L 2016 On cross-frequency phase-phase coupling between theta and gamma oscillations in the hippocampus ed F K Skinner *Elife* **5** e20515 https://doi.org/10.7554/eLife.20515

Sotiriadis P P 2010 Theory of Flying-Adder Frequency Synthesizers—Part I: Modeling, Signals' Periods and Output Average Frequency *Circuits and Systems I: Regular Papers, IEEE Transactions on* **57** 1935–48 http://dx.doi.org/10.1109/TCSI.2009.2039834

Velichko A, Belyaev M, Putrolaynen V, et al., 2018 Thermal coupling and effect of subharmonic synchronization in a system of two $VO_2$ based oscillators *Solid State Electron* **141** 40–9 http://dx.doi.org/10.1016/j.sse.2017.12.003

Velichko A, Belyaev M, Putrolaynen V, et al., 2018 Modeling of thermal coupling in VO<inf>2</inf>-based oscillatory neural networks *Solid State Electron* **139** 8–14 https://doi.org/10.3390/electronics8010075

Velichko A 2019 A method for evaluating chimeric synchronization of coupled oscillators and its application for creating a neural network information converter *Electronics (Switzerland)* **8** https://doi.org/10.3390/electronics8070756

Wang T and Roychowdhury J 2019 OIM: Oscillator-Based Ising Machines for Solving Combinatorial Optimisation Problems pp 232–56 http://dx.doi.org/10.1007/978-3-030-19311-9_19

Xu X, Zheng C and Zhang T 2013 Reduction in LFP cross-frequency coupling between theta and gamma rhythms associated with impaired STP and LTP in a rat model of brain ischemia *Front Comput Neurosci* **7** 27 https://doi.org/10.3389/fncom.2013.00027, PMID: 23576981

Zheng C, Bieri K W, Hsiao Y-T, et al., 2016 Spatial Sequence Coding Differs during Slow and Fast Gamma Rhythms in the Hippocampus *Neuron* **89** 398–408 https://doi.org/10.1016/j.neuron.2015.12.005, PMID: 26774162